\documentclass[12pt]{iopart}

\def\gsim{\raise0.3ex\hbox{$>$}\kern-0.75em{\lower0.65ex\hbox{$\sim$}}}
\newcommand{\om}{\Omega_{\rm M}}
\newcommand{\ola}{\Omega_{\rm\Lambda}}
\usepackage{iopams}
\usepackage{epsfig} 
 
\begin{document}

\title{Constraints on intergalactic dust from quasar colours}
\author{Edvard M\"ortsell\footnote{edvard@physto.se} 
        and Ariel Goobar\footnote{ariel@physto.se},}
\address{Department of Physics, Stockholm University, \\
         S--106 91 Stockholm, Sweden}

\begin{abstract}
Colour measurements of quasars are used to constrain the abundance and
properties of intergalactic dust and the related extinction effects on
high-$z$ sources.  For Type Ia supernovae at $z = 1$, we derive an
upper limit on the possible dimming from intergalactic dust of $\Delta
m=0.2$, ruling out the ``grey'' dust scenario as being solely
responsible for the observed faintness of high-$z$ SNIa.
\end{abstract}

%\submitto{\JCAP}

%\maketitle

%==========
\section{Introduction}
Scattering and absorption of light by a homogeneous distribution of
intergalactic large grains giving a nearly achromatic cross-section --
so called grey dust -- has been proposed \cite{aguirre99a,aguirre99b}
as an alternative to the dark energy explanation for the observed
faintness of Type Ia supernovae (SNIa) at $z\sim 0.5$
\cite{perlmutter,riess}

The presence of a significant component of dark energy has been
inferred from complementary cosmological tests, e.g., from recent WMAP
background radiation measurements \cite{wmap} in combination with
limits on the Hubble parameter from the HST Key Project \cite{hstkey}
and large scale structure measurements from the 2dFGRS survey
\cite{2dfgrs}. However, in Ref.~\cite{blanchard} it was demonstrated how the 
recent CMB anisotropy observations can be explained also by an
Einstein-de Sitter cosmology with no dark energy if one allows for a
low Hubble parameter, $H_0\lesssim 50$ km/s/Mpc, the value currently
favoured by gravitational lensing time delay measurements
\cite{kochanek03}. It is further argued that conflict with large scale structure measurements
can be avoided by suppressing cluster scale mass fluctuations assuming
a small but non-negligible hot dark matter component.

Thus, probing the redshift-distance relation through Type Ia SN
observations remains essential for establishing and
exploring the dark energy of the universe. Ruling out alternative 
explanations to the dark energy interpretation 
and minimizing systematic effects on the measured distances to high-$z$
supernovae is still crucial. 
In this note, we discuss
the possibility to constrain the effects of intergalactic grey dust.

It has been argued that the amount of intergalactic dust required to
make the SNIa results compatible with an Einstein-de Sitter universe
is disfavored by the far-infrared background measured by the
DIRBE/FIRAS instruments \cite{aguirrehaiman}. However, the derived
bounds rely on the assumption that optical and UV photons are absorbed
and re-emitted rather than just scattered by the dust grains. 

It is expected that large dust grains will also cause some differential
extinction for high-$z$ sources (although less than for smaller dust
grains), allowing the possibility to distinguish between extinction
and a cosmological origin of the dimming. In Ref.~\cite{dustcolor}, it
was shown that with 1\,\% relative spectrophotometric accuracy or
broadband photometry in the wavelength interval 0.7--1.5 $\mu$m for
high-$z$ SNIa observations, one could reduce the uncertainty from grey dust
extinction down to $\Delta m=0.02$ magnitudes (the target error for
the proposed SNAP SN survey
\cite{snapprop}).

In this paper, the differential extinction for high-$z$ quasars (QSOs)
caused by extragalactic dust along the line of sight is used to put an
upper limit on the possible effects on SNIa observations at $z = 1$ of
$\Delta m=0.2$ (comparable to current SNIa survey errors) for a wide
range of models for evolution of dust density and extinction
properties.

%==========
%\section{Dust model}
%The most important types of intergalactic grains are silicates and
%graphite. Following Aguirre \cite{aguirre99a,aguirre99b}, we assume
%that intergalactic dust can be described by a Draine \& Lee model
%\cite{drainelee} where the smaller grains have been destroyed by some
%unspecified process, plausibly connected to the expulsion of the dust
%from the star-forming galaxies where it originated.

%The optical properties depend on the value chosen for the small-size
%cutoff $a_{\rm min}$ in the power-law size distribution. The presence
%of small dust grains enhances the wavelength dependence of extinction
%corresponding to lower values of the reddening parameter $R_V$ defined
%by $$ A_V=R_VE(B-V),$$ where $A_V$ is the wavelength-dependent
%extinction and $$E(B-V)=(B-V)-(B-V)_i$$ with $(B-V)_i$ being the
%intrinsic (unobscured) colour.

%==========
\section{Extinction by dust at cosmological distances} 
For a given emission redshift $z_e$, the attenuation $\Delta m_{\rm
d}$ at observed wavelength $\lambda_o$ due to dust can be written
(see also Ref.~\cite{dustcolor})
\begin{equation}
\Delta m_{\rm d}(z_e,\lambda_o)=\frac{2.5}{\ln 10}\int_0^{z_e}
a[\lambda_o/(1+z),R_V]h(z)/D_V(z) dz,\label{eq:deltam}
\end{equation}
where $a(\lambda,R_V)$ is the dimensionless wavelength-dependent
attenuation coefficient where we use the parameterization given in
\cite{ccm} and $D_V(z)\propto [\sigma \cdot \rho_{\rm dust}(z)]^{-1}$
is the interaction length for scattering or absorption by dust
particles in the observed $V$-band ($\lambda =0.55\,\mu$m). Here,
$\rho_{\rm dust}(z)$ is the physical dust density at redshift $z$,
$\sigma$ the interaction cross-section and the proportionality
constant will be related to the mass-distribution of the dust
particles (since $D_V(z)=[\sigma \cdot n_{\rm dust}(z)]^{-1}$ where
$n_{\rm dust}(z)$ is the dust number density). The cosmology-dependent
function $h(z)$ is given by
\begin{equation}
h(z)={1\over H_0\left(1+z\right)\sqrt{(1+z)^2(1+\Omega_{\rm
M}z)-z(2+z)\Omega_\Lambda}}.
\end{equation}
Note that we have set $c=1$ and $h(z)$ and $D_V(z)$ should be given
using the same set of units, e.g., Mpc.

Thus, our parameters are the reddening parameter $R_V$ and the
interaction length at zero redshift $D_{0V} \propto (\sigma
\cdot\rho^\circ_{\rm dust})^{-1}$. 

%\begin{figure}
%  \centerline{\hbox{\epsfig{figure=dustdist,width=.8\textwidth}}}
%  \caption{The magnitude shift (upper panel) and comoving dust-density
%  (lower panel) required to mimic the ``concordance'' cosmology
%  [$\om=0.3$, $\ola=0.7$] for the case of a Einstein-de Sitter cosmology
%  ]$\om=1, \ola=0$]. The dust density is given in units of
%  [$H_0 \cdot m/\sigma$] where $H_0$ is the Hubble parameter, $m$ is the
%  dust particle mass and $\sigma$ is the interaction cross-section.}
%\label{fig:dustdist}
%\end{figure}

The Monte-Carlo simulation program SNOC \cite{snoc} was used to
perform the integral in Eq.~(\ref{eq:deltam}) numerically. Note that
the model is valid also for a patchy dust distribution, as long as the
scale of inhomogeneities is small enough, i.e., $1/\sqrt{N}\ll 1$
where $N$ is the number of dust clouds intersected by the light-ray.

%In Fig.~\ref{fig:dustdist}, the magnitude shift (upper panel) and
%comoving dust-density (lower panel) required to mimic the
%``concordance'' cosmology [$\om=0.3$, $\ola=0.7$] for the case of a
%Einstein-de Sitter cosmology [$\om=1, \ola=0$] is shown.  
Perfectly grey dust (i.e., with wavelength-independent absorption) can
obviously mimic the effects of any dark energy component if one
allows for arbitrary fine-tuning of the dust distribution. We have
investigated whether any realistic dust model related to astrophysical
sources are consistent with the available data. We note first that the dust density has
to be related to astrophysical sources, such as star formation, and
second, that physically reasonable dust models generally implies a
wavelength-dependence in the absorption and scattering properties.

As in Ref.~\cite{dustcolor}, two dust distributions are considered:
$\rho_{\rm dust}\propto (1+z)^\alpha$, where
\begin{eqnarray}\label{eq:dustdensity}
 \alpha(z) =  
      \left\{ 
        \begin{array}{lc}
          3 {\rm \ for \ all \ }z,&  {\rm \ Model \ A}\\ 
          0 {\rm \ for \ } z>0.5 {\rm \ (3 \ for \ lower \ } z). &  {\rm \ Model \ B} 
       \end{array}
        \right.   
\end{eqnarray}

Thus, Model A has a constant comoving dust density whereas Model B
with $\rho_{\rm dust}(z>0.5) = \rho_{\rm dust}(z=0.5)$ implies dust
creation at $z>0.5$. Models A and B are considered to be extreme in
opposite directions, i.e. they should capture the full range of
physical models leading to the production of intergalactic dust.

In our simulations we explore values of $R_V$ in the interval ranging
from 1 to 9 and the interaction scale-length at zero redshift $D_{0V}$
in the range $10-10\,000$ Gpc. In a $[\om =0.3,\ola=0.7]$-cosmology,
$z=[0.5, 1.0, 1.5]$ correspond to proper distances of 1.55, 2.37 and
2.84 Gpc, respectively.  Note that since the dust density varies with
redshift according to Eq.~(\ref{eq:dustdensity}), the interaction
length will vary correspondingly.

%==========
\section{Spectral homogeneity of QSOs}
QSOs have been found to have remarkably homogeneous spectral
properties in spite of their wide range in brightness. The Sloan
Digital Sky Survey collaboration (SDSS, \cite{sdss}) has produced
empirical spectral templates based on a couple of thousand objects in
the redshift range $0.044<z<4.789$ showing a spectrum-to-spectrum
$1\,\sigma$ difference of approximately 20\,\% in the rest system
wavelength range 1200 to 8500 \AA \ \cite{sdsscomp}.  HST QSO data was
used by \cite{telfer} to extend the spectral template down to 300 {\AA} 
in order to obtain full wavelength coverage for all broadband filters
(see Sec.~\ref{sec:sample}) over the entire redshift interval used in
this paper.

%==========
\section{The QSO sample}\label{sec:sample}
\begin{figure}
  \centerline{\hbox{\epsfig{figure=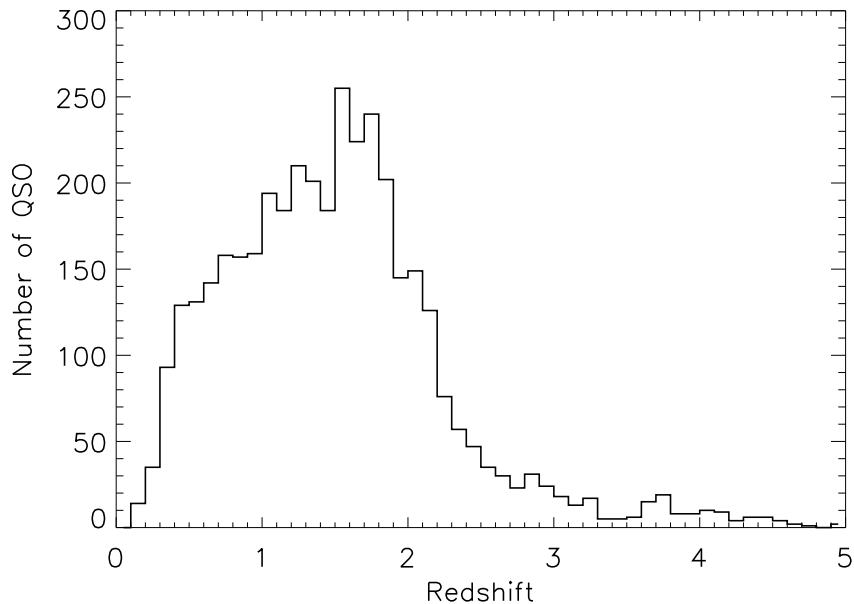,width=.8\textwidth}}}
  \caption{Redshift distribution of 3814 QSOs in the Sloan Digital Sky
  Survey Early Data Release.}  
\label{fig:zdistr}
\end{figure}

The SDSS and 2dF \cite{2df} Galaxy Redshift Survey will eventually
include about 10$^5$ QSOs up to redshift $z\sim 5$.  We have used a
data sample of 3814 QSOs from the SDSS Early Data Release (EDR)
\cite{sdssedr}. These have redshifts between $z=0.15$ and $5.03$ distributed
according to Fig.~\ref{fig:zdistr}.

%The survey uses a dedicated 2.5 m telescope, located at Apache Point
%Observatory (APO) in New Mexico, with a 3 degree field of view. The
%photometry data was calibrated with the aid of an auxiliary 0.5 m
%photometric telescope (PT).

The optical magnitudes are measured through five broadband filters
($ugriz$) with limiting point-source magnitudes of 22.3, 22.6, 22.7,
22.4, and 20.5, respectively, for a signal-to-noise ratio of 5:1.

%The photometric calibration is somewhat uncertain since the filters
%used on the PT differ systematically from those on the 2.5 m camera,
%and one does not yet have a complete understanding of the
%transformations between the systems. 
The quoted photometric calibration errors are 3\,\% in $g, r$ and $i$,
and 5\,\% in $u$ and $z$ \cite{sdssedr}. The instrumental response
functions are given in Ref.~\cite{respt} where we use the second
column ({\tt respt}) corresponding to 1.3 airmasses at APO for a point
source.

We exclude QSOs that are classified as extended to avoid contamination
from the host galaxy. This cut leaves 3692 of the original 3814 QSOs.
Applying the 5:1 signal-to-noise ratio limiting point-source magnitude
cut leaves 3556 of the point-like QSOs. In order to avoid confusion
with Seyfert galaxies, we also exclude objects at $z<0.5$
\cite{richards1,richards2}. To avoid putting 
excessive weight in the analysis on a possible dust distribution at
very high redshifts (and also to minimize a possible bias from flux
limitations) we have only considered QSOs with redshift $z<2$ in our
primary analysis, leaving 2740 point-like QSOs at $0.5<z<2$. An
alternative method to constrain intergalactic dust at $z>2$ by
studying the thermal history of the intergalactic medium is presented
in Ref.~\cite{inoue}.

%==========
\section{Method and results}\label{sec:method}
We divide the QSO sample into redshift bins of size $\Delta z =
0.1$. In each redshift bin, we calculate the mean colour, e.g.,
$(u-g)_{\rm obs}$ as a measurement error weighted average. Objects
that have colours that are more than $2\,\sigma$ off from the mean
value are rejected to avoid contamination from object with anomalous
colours, e.g., QSOs that have suffered severe extinction in the host
galaxy.
%After this final cut in our original sample of 3814 QSO, there remains
%3231 QSOs.   

For each set of parameter values $R_V=[1,9],D_{0V}=[10,5\,000]$ Gpc and
redshift, we simulate the attenuation due to grey dust and add to the
median QSO spectrum obtained in Ref.~\cite{sdsscomp}.

For a pair of broadband filters $X$ and $Y$,
we define 
\begin{equation}
\Delta(X-Y)_{z,R_V,D_{0V}}=(X-Y)_z-({\cal X-Y})_{z,R_V,D_{0V}}
\end{equation}
where $(X-Y)_z$ is the observed colour of a QSO at redshift $z$ and
$({\cal X-Y})_{z,R_V,D_{0V}}$ the simulated colour for a specific set
of dust parameter values, i.e, the colour from the template spectrum
corrected for differential extinction and convolved with the SDSS
filter functions. We then calculate the $\chi^2$ as a function of
$R_V$ and $D_{0V}$ as
\begin{equation}
\chi^2[R_V,D_{0V}]=\sum_{i,j=1}^N \Delta(X-Y)_{z_i,R_V,D_{0V}}(V(X-Y)^{-1})_{i,j}\Delta(X-Y)_{z_j,R_V,D_{0V}}
\end{equation}
where the sum is over $N$ redshift bins and $V(X-Y)_{i,j}$ is the
covariance matrix defined as
\begin{equation}
V(X-Y)_{i,j}=\sigma(X)^2_{\rm sys}+\sigma(Y)^2_{\rm sys}+\delta_{ij}\left[
\sigma(X)^2_{\rm tem}+\sigma(Y)^2_{\rm tem}+\sigma(X-Y)_{\rm obs}^2\right]
\end{equation}

Here, $\sigma_{\rm sys}$ denote the systematic errors associated with
each broadband filter (see Sec.~\ref{sec:sample}), $\sigma_{\rm tem}$
the template error (Sec.~\ref{sec:evolution}) and $\sigma_{\rm obs}$
the error in the mean observed colour in each redshift bin (after the
cut of outliers described above). The scatter in each redshift
bin is approximately 0.2 mag so with the order of 100 QSOs in each
redshift bin we have $\sigma_{\rm obs}\sim 0.02$ mag.
 
\begin{figure}
  \centerline{\hbox{\psfig{figure=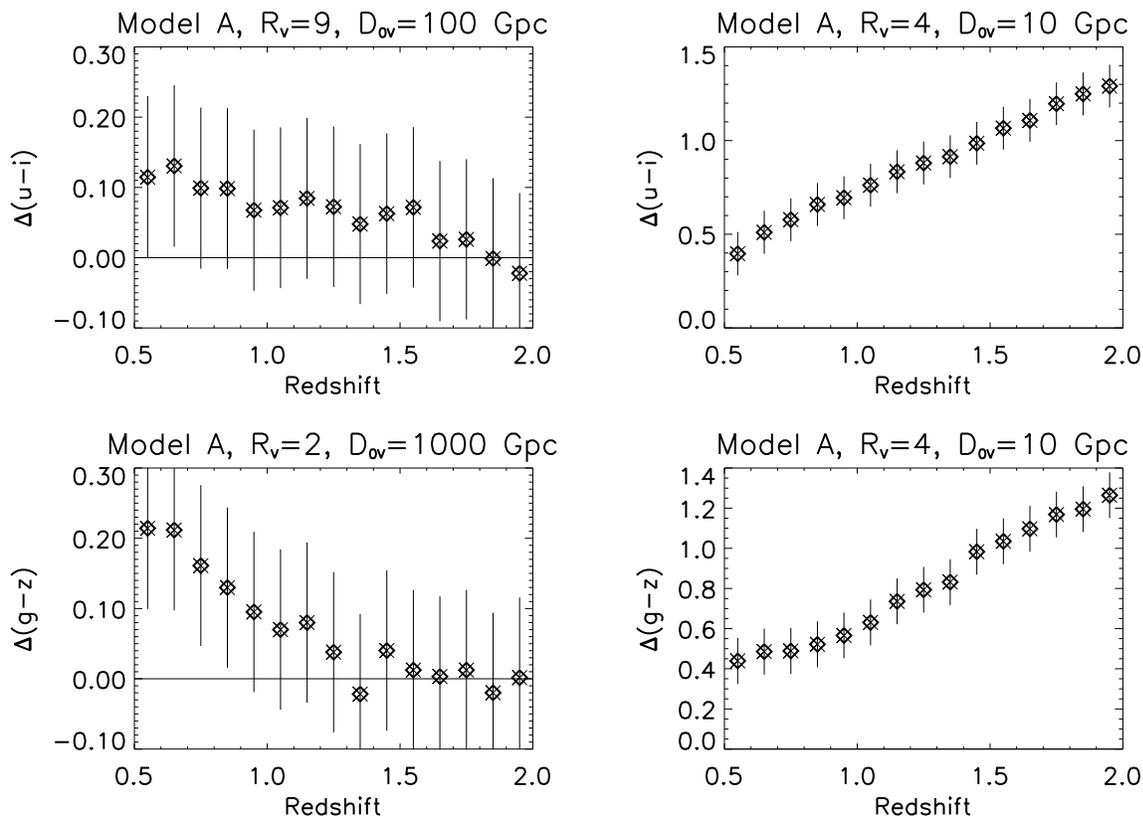,width=1.\textwidth}}}
  \caption{The colour-redshift evolution for different dust
  scenarios. The upper left panel shows $\Delta (u-i)$ vs $z$ for $R_V = 9$
  and $D_{0V} = 100$ Gpc (Model A), the lower left panel shows $\Delta (g-z)$ vs
  $z$ for $R_V = 2$ and $D_{0V} = 1000$ Gpc (Model A). The right panels
  show corresponding plots for $R_V = 4$ and $D_{0V} = 10$ Gpc that give
  very low probabilities for the same dust distribution model.}
\label{fig:uigz}
\end{figure}

In Fig.~\ref{fig:uigz}, the colour-redshift evolution for a couple of
different dust scenarios are compared. The upper left panel shows
$\Delta(u-i)$ vs redshift for the Model A best fit parameters $R_V =
9$ and $D_{0V} = 100$ Gpc and the lower left panel shows $\Delta(u-i)$
vs redshift for the Model A best fit parameters $R_V = 2$ and $D_{0V}
= 1000$ Gpc. These parameters yield $\chi^2/{\rm dof}\sim 0.5$,
indicating that systematic errors may have been overestimated (see
Sec.~\ref{sec:sys}). These plots are to be compared with the case of
$R_V = 4$ and $D_{0V} = 10$ Gpc that give high $\chi^2$-values (i.e.,
low probabilities) for the same dust distribution model (right
panels).

\begin{figure}
  \centerline{\hbox{\psfig{figure=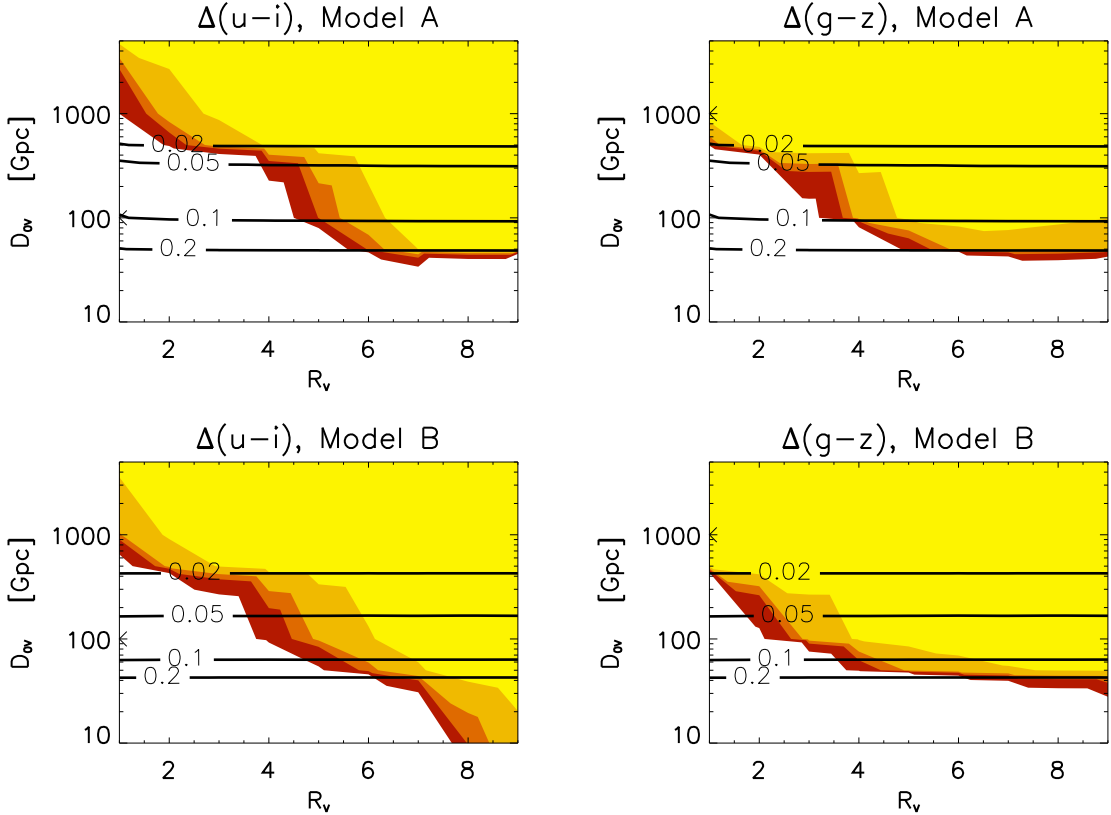,width=1.\textwidth}}}
  \caption{Results from restframe $B$-band attenuation for SNIa
  at $z=1$ and the $\chi^2$-test using $\Delta (u-i)$ (left panels) and $\Delta (g-z)$
  (right panels). Contour levels labeled [0.2,0.1,0.05,0.02] show the
  amount of attenuation (in magnitudes) while the red, dark orange,
  light orange, and yellow regions indicate 68, 90, 95 and 99\,\%
  confidence level allowed regions from the $\chi^2$-test. Small scale
  features are due to low resolution in the parameter grid.}
\label{fig:uigzsep}
\end{figure}

We have also investigated the dimming effect on high-$z$ sources at
different wavelengths, in particular the restframe $B$-band flux of
SNIa. In Fig.~\ref{fig:uigzsep}, we have combined the results from
this investigation with the results from the $\chi^2$-test. In the
left panels, the results for $\Delta(u-i)$ assuming the Model A (upper
panel) and Model B (lower panel) dust distribution are shown. The
right panels show the corresponding results for $\Delta(g-z)$.
%Note that this is a {\em very} conservative estimation since we have
%not taken into consideration the intrinsic variation of QSO spectra.
The red, dark orange, light orange, and yellow regions indicate 68,
90, 95 and 99\,\% confidence level allowed regions from the
$\chi^2$-test. The four contour levels labeled [0.2,0.1,0.05,0.02]
correspond to the amount of attenuation caused by grey dust when
integrating over the restframe $B$-band of a set of SNIa at a
redshift of $z=1$.

\begin{figure}
  \centerline{\hbox{\psfig{figure=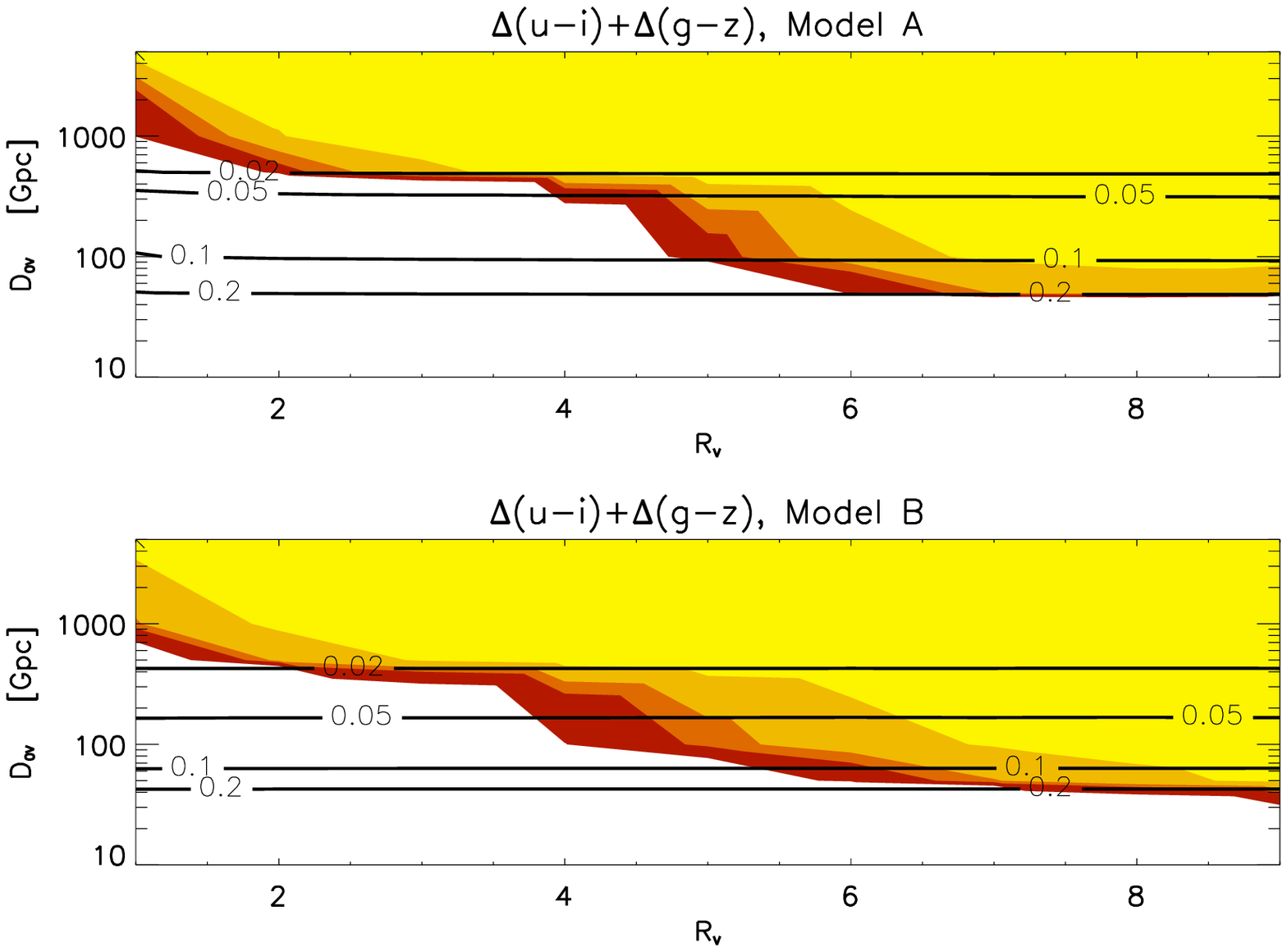,width=1.\textwidth}}}
  \caption{Combined results from restframe $B$-band attenuation for SNIa
  at $z=1$ and the $\chi^2$-test using $\Delta (u-i)$ and $\Delta (g-z)$. Contour
  levels labeled [0.2,0.1,0.05,0.02] show the amount of attenuation
  (in magnitudes) while the red, dark orange, light orange, and yellow
  regions indicate 68, 90, 95 and 99\,\% confidence level allowed
  regions from the $\chi^2$-test.}
\label{fig:uigzcontour}
\end{figure}

In Fig.~\ref{fig:uigzcontour}, we show combined results from colours
$\Delta (u-i)$ and $\Delta (g-z)$. Allowing for any value of $R_V$ within the
parameter range, we are able to rule out $\Delta m>0.2$ at 99\,\%
confidence level for SNIa at $z=1$. Assuming intergalactic dust
properties more similar to galactic dust ($R_V
\lesssim 4$) we are able to do almost a factor of ten better ($\Delta
m\lesssim 0.03$). Note also that all our results are consistent with a
no dust hypothesis since the $\chi^2$-values flattens out when
$D_{0V}\to\infty$. The small scale features in Fig.~\ref{fig:uigzsep}
and \ref{fig:uigzcontour} are due to low resolution in the parameter
grid.

Redoing the analysis combining $\Delta (u-z)$ and $\Delta (g-i)$ yields very
similar results. 

%Including also QSO data at $z>2$ allows us to put
%very strong limits ($\Delta m \lesssim 0.02$) in the case of comoving
%dust density (Model A). For model B, results are not very sensitive to
%the inclusion of very high-$z$ QSOs.

%==========
\section{Systematic effects}\label{sec:sys}

\subsection{Selection effects}
The selection algorithm for the SDSS quasar sample varied throughout the
time that the EDR observations were obtained, so the objects in this
catalog were not found via a uniform set of selection criteria
\cite{schneider}. Richards et al \cite{richards3} analysed the
limitations on the colour range of QSOs and found that for $z=2$, the
SDSS survey was sensitive to reddened QSOs with $E(B-V)\gtrsim
0.54$. We conclude that selection effects do not significantly affect
our ability to compute the average colour of the QSO sample out to
redshift $z=2$.

\subsection{Spectral evolution of QSOs}\label{sec:evolution}
The possible spectral evolution is limited by the nearly Gaussian
20\,\% variations found for the whole sample. As we only consider less
than half of the survey's redshift range, systematic colour trends
with redshift would not exceed $\sim$ 10\,\%. In our analysis, we use
a systematic uncertainty on the magnitude derived from the median
spectrum vs redshift of 10\,\% in $u$ and $z$ and 5\,\% in $g,r$ and
$i$. In general, unless the spectral evolution trends exactly cancel
the differential extinction from dust, our limits on grey dust are
conservative in the sense that the reddening that we attribute to
possible extinction by dust may stem from a redder population of QSOs.
Of course, a possible problem is the fact that the template spectrum
we use is derived from spectra that themselves might be affected by
dust absorption and extinction. Such an effect would not cause any
major change in differential evolution of the QSO colours vs
redshift. However, it should be noted that it could cause a close to
constant offset in the colour-redshift diagram that might bias the
outcome of the $\chi^2$-analysis. An iterative method where this is
corrected for by shifting the colour of the template to match the
colour of the current best fit model at the mean redshift of the QSO
sample ($z\sim 1.4$) yields results similar to
Fig.~\ref{fig:uigzcontour} with the exception that for Model B, the
upper right corner (corresponding to high values of $R_V$ and
$D_{0V}$) is also excluded. However, the main conclusion that $\Delta
m>0.2$ is excluded for SNIa at $z=1$ remains unchanged. Improved
statistics and systematic errors in the QSO surveys will improve the
control of systematic effects from the template spectrum, see
Sec.~\ref{sec:future}.

%We generally get rather low $\chi^2_{\rm min}$ values, e.g., for
%$\Delta (u-i)$ we have $\chi^2_{\rm min}/{\rm ndf}\sim 0.3$ and for $\Delta (g-z)$
%we have $\chi^2_{\rm min}/{\rm ndf}\sim 0.6$ indicating that if
%anything, systematic errors have been overestimated.
 
\subsection{Host galaxy dust}
Internal reddening may come from depletion by material associated with
the torus of gas and dust that is thought to surround the central QSO
engine \cite{sdssedr}. Richards et al. \cite{richards3} estimate that
6\,\% of the SDSS quasar sample shows signs of reddening by dust,
probably in the host environment. The computation of average colours
for each redshift bin takes this into account by an iterative
rejection mechanism where QSOs more than $2\,\sigma$ away from the
average colour at that redshift are rejected, see
Sec.~\ref{sec:method}. Thus, the tails of the distribution are
suppressed. Remaining reddened QSOs would tend to make our limits more
conservative, as the residual host galaxy dust would be counted as
possible intergalactic dust.

\subsection{Evolution of dust properties}
Although we have assumed a constant value of $R_V$ for the redshift
range $0.5 < z < 2.0$, our derived limits show only moderate
dependence on the actual value of the reddening parameter. A variation
of $R_V$ within the considered range would thus not lead to a
significant difference in the observed attenuation of high-$z$ SNIa.

%==========
\section{Future prospects}\label{sec:future}
The Sloan Digital Sky Survey has recently made available the Data
Release One (DR1) with spectroscopic parameters for 17\,700 medium-$z$
($z<2.3$) and 980 high-$z$ QSOs ($z>2.3$), including reprocessed data
from the Early Data Release. A catalog of DR1 QSOs is being prepared
\cite{dr1} and we plan to redo the analysis with the full data set when
the catalog is completed. The increased number of QSOs will of course
help in decreasing the statistical error. Hopefully, the added data
will also be helpful in order to improve our understanding of
different systematic effects. For example, it should be possible to
constrain the possible evolution of QSO colours and minimize the
systematic error from the spectral template by studying the
homogeneity of QSO spectra over a range of redshifts.

%==========
\section{Summary and conclusions}
Studying the redshift evolution of colours of QSOs from the Sloan
Digital Sky Survey Early Data Release, we constrain the effect from
extinction by homogeneous intergalactic dust on high-$z$
observations. We compare observed spectra from 2740 QSOs at $0.5<z<2$
with simulated spectra obtained by adding the attenuation due to
different grey dust scenarios to a reference median QSO spectrum. 
For each dust model, we also compute the related extinction effects on
high-$z$ SNIa.

For a wide range of models for the dust distribution and extinction
properties, we are able to rule out $\Delta m\gtrsim 0.2$ at 99\,\%
confidence level for SNIa at $z=1$. This is comparable to current SNIa
survey errors and allows us to rule out the ``grey'' dust scenario as
an alternative to the cosmological constant explanation for the
observed faintness of high-$z$ SNIa. It should be noted that these
results are based on the specific parameterization of the
wavelength-dependent dust attenuation given in Ref.~\cite{ccm} and
that the validity of this specific parameterization for very high
values of $R_V$ remains untested.

To constrain the dark energy properties (e.g., the equation of state
parameter) with future SN surveys such as SNAP, yet better constrains
on dimming by grey dust are required.  Assuming intergalactic dust
properties similar to galactic dust ($R_V \lesssim 4$) the present
analysis rules out $\Delta m\gtrsim 0.03$. However, for a more general dust
scenario, larger statistical samples and better control of the
systematics of the QSO sample are required. Hopefully this will be
within reach with future data sets from, e.g., the Sloan Digital Sky
Survey.

%==========
\section*{Acknowledgments}

The authors would like to thank Anthony Aguirre and the anonymous
referee for careful reading and valuable suggestions improving the
quality of the manuscript. They would also like to thank Mamoru Doi
and Gordon Richards for help regarding the SDSS response
functions. A.G. is a Royal Swedish Academy Research Fellow supported
by a grant from the Knut and Alice Wallenberg Foundation.

Funding for the creation and distribution of the SDSS Archive has been
provided by the Alfred P.  Sloan Foundation, the Participating
Institutions, the National Aeronautics and Space Administration, the
National Science Foundation, the US Department of Energy, the Japanese
Monbukagakusho, and the Max Planck Society. The SDSS Web site is
http://www.sdss.org/. The Participating Institutions are the
University of Chicago, Fermilab, the Institute for Advanced Study, the
Japan Participation Group, the Johns Hopkins University, the Max
Planck Institute for Astronomy (MPIA), the Max Planck Institute for
Astrophysics (MPA), New Mexico State University, Princeton University,
the United States Naval Observatory, and the University of Washington.

%========== 
\end{document}